
%
\input harvmac

\def\({[}
\def\){]}
\def\hqs{{\hat q}^2}
\def\hmts{\rho_\tau}
\def\rt{\rho_\tau}
\def\hz{{\hat z}}
\def\r{\rho_j}
\def\Mt{\hat m_\tau}
\def\lo{\lambda_1}
\def\lt{\lambda_2}

\def\gev{{\;\rm GeV}}
\def\lambar{{\overline\Lambda}}

\noblackbox

\Title{\vbox{
\hbox{CERN-TH.7124/93}
\hbox{JHU-TIPAC-930031}
\hbox{UCSD/PTH 93-43}
\hbox{WIS-93/117/Dec-PH}}}
{\vbox{
\centerline{Heavy Quark Expansion for the}
\bigskip
\centerline{Inclusive Decay $\bar B\to\tau\,\bar\nu\,X$}}}
\smallskip
\centerline{Adam F. Falk$^{ab}$, Zoltan Ligeti$^c$,
Matthias Neubert$^d$, and Yosef Nir$^c$}
\bigskip
\centerline{\it $^a$Department of Physics, University of California,
San Diego}
\centerline{\it La Jolla, California 92093, USA}
\smallskip
\centerline{\it $^b$Department of Physics and Astronomy, The Johns
Hopkins University}
\centerline{\it Baltimore, Maryland 21218, USA}
\smallskip
\centerline{\it $^c$Department of Particle Physics}
\centerline{\it Weizmann Institute of Science, Rehovot 76100, Israel}
\smallskip
\centerline{\it $^d$Theory Division, CERN}
\centerline{\it CH-1211 Geneva 23, Switzerland}
\bigskip
\baselineskip 18pt

\centerline{\bf Abstract:}
\medskip
\noindent
We calculate the differential decay rate for inclusive $\bar
B\to\tau\,\bar\nu\,X$ transitions to order $1/m_b^2$ in the heavy
quark expansion, for both polarized and unpolarized tau leptons. We
show that using a systematic $1/m_b$ expansion significantly reduces
the theoretical uncertainties in the calculation. We obtain for the
total branching ratio ${\rm BR}(\bar B\to\tau\,\bar\nu\,X)=2.30\pm
0.25\%$, and for the tau polarization $A_{\rm pol}=-0.706\pm0.006$.
{}From the experimental measurement of the branching ratio at LEP, we
derive the upper bound $\lo\leq 0.8\gev^2$ for one of the parameters
of the heavy quark effective theory.

\Date{December 1993}

\newsec{Introduction}

Recently, it has been observed that inclusive semileptonic decays of
hadrons containing a single heavy quark allow for a systematic,
QCD-based expansion in powers of $1/m_Q$~\ref\cgg{J. Chay, H. Georgi,
and B. Grinstein, Phys.\ Lett.\ B247 (1990) 399.}. In particular, it
has been shown that the inclusive decay rates computed in the heavy
quark limit $m_Q\to\infty$ coincide with those obtained in the free
quark decay model, while corrections of order $1/m_Q$ vanish. The
leading nonperturbative corrections are of order $1/m_Q^2$ and depend
on only two hadronic parameters, which parameterize certain forward
matrix elements of local dimension-five operators. These corrections
have been computed for a number of processes~\ref\bsuv{I.I. Bigi, M.
Shifman, N.G. Uraltsev, and A. Vainshtein, Phys.\ Rev.\ Lett.\ 71
(1993) 496.}--\nref\bksv{B. Blok, L. Koyrakh, M. Shifman, and A.I.
Vainshtein, NSF-ITP-93068 (1993), hep-ph/9307247.}\nref\mawi{A.V.
Manohar and M.B. Wise, UCSD/PTH 93-14 (1993),
hep-ph/9308246.}\nref\fls{A.F. Falk, M. Luke, and M.J. Savage,
UCSD/PTH 93-23 (1993), hep-ph/9308288, to appear in Phys.\ Rev.\
D.}\nref\mann{T. Mannel, IKDA 93/26 (1993),
hep-ph/9309262.}\ref\neub{M. Neubert, CERN-TH.7087/93 (1993),
hep-ph/9311325, to appear in Phys.\ Rev.\ D; CERN-TH.7113/93 (1993),
hep-ph/9312311.}. These new theoretical developments not only provide
a theoretical justification for the parton model, but also allow a
model independent calculation of the nonperturbative corrections to a
high level of accuracy.

Semileptonic $B$ meson decays into a tau lepton are potentially very
interesting. First, they are sensitive to certain form factors which
are unmeasurable with a massless lepton in the final state. As a
consequence, the ratio of the decay rates for the tau channel and the
light lepton channels is sensitive to the nonperturbative corrections
of order $1/m_b^2$, while independent of the
Cabbibo--Kobayashi--Maskawa matrix elements $V_{cb}$ and $V_{ub}$.
Second, the possibility to study the tau polarization offers a
greater variety in the experimental and theoretical analysis and an
independent determination of parameters. Third, the inclusive $\bar
B\to\tau\,\bar\nu\,X$ decay rate is useful for constraining or
probing certain extensions of the standard model, such as models with
many scalar fields.

In this paper, we study the decay $\bar B\to\tau\,\bar\nu\,X$ using a
combination of the operator product expansion and heavy quark
effective theory. The main points which we address are the following:

$a$. We calculate analytically, to order $1/m_b^2$, the total decay
rate and the lepton spectrum for the cases of unpolarized and
polarized tau leptons. Effects of the finite tau lepton mass are
treated exactly.

$b$. We study in detail the numerical predictions for the total decay
rate and the tau polarization. In particular, we stress that by using
a systematic $1/m_Q$ expansion the theoretical predictions become
more accurate than those of the free quark decay model, not only
because $1/m_b^2$ corrections are included but, more importantly,
because the masses of the charm and bottom quarks are correlated in a
specific way.

$c$. We use the experimental value of the inclusive branching ratio
${\rm BR}(\bar B\to\tau\,\bar\nu\,X)$ to derive a bound on the
hadronic parameter $\lambda_1$, which is related to the kinetic
energy of the $b$-quark inside the $B$ meson.

\newsec{Analytic Expressions}

We begin by presenting the analytic results necessary for our
numerical analysis. The techniques by which they are obtained are
described in detail elsewhere \bsuv--\fls, so we will present only a
brief summary, followed by the results of the computation.

The inclusive differential decay distribution is determined by the
imaginary part of the time-ordered product of two flavor-changing
currents,
\eqn\timeordered{
     T^{\mu\nu}=-i\int d^4x\,e^{-iq\cdot x}\,\langle B|\,
     T\left\{J^{\mu\dagger}(x),J^\nu(0)\right\} |B\rangle\,,}
where $J^\mu=\overline c\,\gamma^\mu(1-\gamma_5)\,b$ or $J^\mu=
\overline u\,\gamma^\mu(1-\gamma_5)\,b$. Since over most of the
Dalitz plot the energy release is large (of order $m_b$), the
time-ordered product can be written as an operator product expansion,
in which higher-dimension operators are suppressed by powers of
$\Lambda/m_b$, where $\Lambda$ is a typical low energy scale of the
strong interactions. To this end, however, it is necessary to
separate the large part of the $b$-quark momentum by writing $p_b=m_b
v+k$, where $v$ is the velocity of the decaying $B$ meson. The aim is
to construct an expansion in powers of $k/m_b$, where the residual
momentum $k$ is of order $\Lambda_{\rm QCD}$. This separation is most
conveniently performed by using the formalism of the heavy quark
effective theory~\ref\HQET{For a review, see: M. Neubert,
SLAC-PUB-6263 (1993), to appear in Phys.\ Rep., and references
therein.}. One can then evaluate the matrix elements of the resulting
tower of nonrenormalizable operators with the help of the heavy quark
symmetries.

The leading term in the expansion reproduces the result of the free
quark decay model \cgg, while giving a unambiguous meaning to the
heavy quark mass~\ref\AMM{A.F. Falk, M. Neubert, and M. Luke, Nucl.\
Phys.\ B388 (1992) 363; M. Neubert, Phys.\ Rev.\ D46 (1992) 3914.}.
The leading nonperturbative corrections are of relative order
$1/m_b^2$ and may be written in terms of two parameters, $\lambda_1$
and $\lambda_2$, which are related to the kinetic energy $K_b$ of the
$b$-quark inside the $B$ meson, and to the mass splitting between $B$
and $B^*$ mesons~\ref\FaNe{ A.F. Falk and M. Neubert, Phys.\ Rev.\
D47 (1993) 2965.}:
\eqn\deflambda{
   K_b = - {\lambda_1\over 2 m_b}\,,\qquad
   m_{B^*}^2 - m_B^2 = 4\lambda_2\,.}

The operator product expansion for semileptonic decays to massless
leptons has been performed in refs.~\bsuv--\mawi. The inclusion of
the tau mass in the process $\bar B\to\tau\,\bar\nu\,X$ is a
straightforward, but cumbersome, generalization. For the sake of
brevity, we shall present only the final expressions. The tau lepton
can have spin up ($s=+$) or spin down ($s=-$) relative to the
direction of its momentum, and it is convenient to decompose the
corresponding decay rates as
\eqn\gamtilgam{\Gamma\left(\bar B\to\tau(s=\pm)\,\bar\nu\,X\right)=
    {1\over2}\Gamma\pm\tilde\Gamma\,.}
The total rate, summed over the tau polarizations, is given by
$\Gamma$, while the tau polarization is $A_{\rm pol}=2\tilde\Gamma/
\Gamma$. The differential decay rate depends on the kinematic
variables $q^2$, $E_\tau$, and $E_\nu$, where $q^2$ is the invariant
mass of the lepton pair, and $E_\tau$ and $E_\nu$ denote the tau and
neutrino energies in the parent rest frame. Let us introduce a set of
related dimensionless variables by
\eqn\defone{
    \hqs={q^2\over m_b^2}\,,\qquad
    y={2E_\tau\over m_b}\,,\qquad
    x={2E_\nu\over m_b}\,,}
and define the mass ratios
\eqn\deftwo{
    \r={m_j^2\over m_b^2}\,,\qquad \hmts={m_\tau^2\over m_b^2}\,,}
where $j=c$ or $u$ is a flavor label. The triple differential decay
rate may be written in terms of five invariant form factors $\hat
W_i$, for which we adopt the conventions of ref.~\mawi. We obtain
\eqn\upoldr{\eqalign{
    {1\over\Gamma_b}
    {d\Gamma\over d\hqs dy dx} &= 24\,\Theta\left(x-
    {2(\hqs-\hmts)\over y+\sqrt{y^2-4\hmts}}\right)
    \Theta\left({2(\hqs-\hmts)\over y-\sqrt{y^2-4\hmts}}-x\right)\cr
    &\times \left\{(\hqs-\rt)\hat W_1+{1\over2}
    (yx-\hqs+\rt)\hat W_2
    +{1\over2}\(\hqs(y-x)-\rt(y+x)\)\hat W_3\right.\cr &\ \ \ \left.+
    {1\over2}\rt(\hqs-\rt)\hat W_4+\rt x\hat W_5\right\}\,,\cr}}
and
\eqn\poldr{\eqalign{{1\over\Gamma_b}
    {d\tilde\Gamma\over d\hqs dy dx} &= 6\,\Theta\left(x-
    {2(\hqs-\hmts)\over y+\sqrt{y^2-4\hmts}}\right)
    \Theta\left({2(\hqs-\hmts)\over y-\sqrt{y^2-4\hmts}}-x\right)
    {1\over\sqrt{y^2-4\rt}}\cr &\times \bigg\{\((\hqs-\rt)y-2\rt x\)
    (-2\hat W_1+\hat W_2+x\hat W_3+\rt\hat W_4+y\hat W_5)\cr
 &\ -(y^2-4\rt)\(x\hat W_2+(\hqs-\rt)(\hat W_3+\hat
W_5)\)\bigg\}\,,\cr}}
where
\eqn\gammab{
    \Gamma_b={|V_{jb}|^2 G_F^2\, m_b^5 \over 192\pi^3} \,;\quad
    j=c\ {\rm or}\ u.}
The expressions for the form factors $\hat W_i$, at tree level and to
order $1/m_b^2$ in the operator product expansion, are
\eqn\wsubi{\eqalign{
  \hat W_1 &= \delta(\hz) \Big\{ {1\over 4}(2-y-x)
    - {\lo+3\lt\over 12 m_b^2} \Big\} \cr
   &+ \delta'(\hz) \Big\{ {\lo\over 24 m_b^2}
    \( 8\hqs+6(y+x)-5(y+x)^2\)
    + {\lt\over 8 m_b^2} \(8\hqs+14(y+x)-5(y+x)^2-16\) \Big\} \cr
   &+ \delta''(\hz) {\lo\over 24 m_b^2}(2-y-x)
    \(4\hqs-(y+x)^2\) \,, \cr
  \hat W_2 &= \delta(\hz) \Big\{ 1-{5(\lo+3\lt)\over 6 m_b^2} \Big\}
    + \delta'(\hz) \Big\{ {7\lo\over 6 m_b^2} (y+x)
    + {\lt\over 2 m_b^2} \(5(y+x)-4\) \Big\} \cr
   &+ \delta''(\hz) {\lo\over 6 m_b^2} \(4\hqs-(y+x)^2\) \,, \cr
  \hat W_3 &= {\delta(\hz)\over 2}
    + \delta'(\hz) \Big\{ {5\lo\over 12 m_b^2}(y+x)
    + {\lt\over 4 m_b^2} \(5(y+x)-12\) \Big\}
    + \delta''(\hz) {\lo\over 12 m_b^2} \(4\hqs-(y+x)^2\) \,, \cr
  \hat W_4 &= \delta'(\hz) {2(\lo+3\lt)\over 3 m_b^2} \,, \cr
  \hat W_5 &= -{\delta(\hz)\over2} - \delta'(\hz) \Big\{
    {\lo\over 12 m_b^2} \(5(y+x)+8\) + {5\lt\over 4 m_b^2}(y+x)
    \Big\} - \delta''(\hz) {\lo\over 12 m_b^2} \(4\hqs-(y+x)^2\)
    \,, \cr}}
where
\eqn\defhz{
    \hz=1+\hqs-\r-y-x+i\epsilon \,.}
Our results for $\hat W_1$, $\hat W_2$ and $\hat W_3$ coincide with
those obtained in ref.~\mawi. The two form factors $\hat W_4$ and
$\hat W_5$ do not contribute when the final lepton is massless, but
are important for the tau channel.

We now integrate over the kinematic variables $\hat q^2$ and $x$ to
obtain the differential decay rates with respect to the rescaled
lepton energy $y$. Notice that a subtlety specific for the case of
non-vanishing tau lepton mass is the appearance of the second step
function in \upoldr\ and \poldr, which gives an upper limit for the
variable $x$. For the case of a massless lepton, this limit becomes
trivial ($x<\infty$). Integrating \upoldr, we find
\eqn\dgammady{\eqalign{
    {1\over\Gamma_b}&{d\Gamma\over dy}=2\sqrt{y^2-4\rt}\bigg\{
    x_0^3\Big\(y^2-3y(1+\rt)+8\rt\Big\)
    +x_0^2\Big\(-3y^2+6y(1+\rt)-12\rt\Big\)\cr
    &-{\lt\ x_0\over m_b^2(1+\rt-y)}\Big\(
    5x_0^2\left(y^3-4y^2(1+\rt)+2y(3+7\rt)+4\rt(\rt-5)\right)\cr
    &\
+3x_0\left(-5y^3+y^2(17+15\rt)-y(24+46\rt)-\rt(18\rt-70)\right)\cr
    &\
+3\left(5y^3-10y^2(1+\rt)+4y(3+4\rt)+16\rt(\rt-2)\right)\Big\)\cr
    &+{\lo\over3m_b^2(1+\rt-y)^2}\Big\(
    3\left(y^4-2y^3(1+\rt)+8y\rt(1+\rt)-16\rt^2\right)\cr
    &\
+2x_0^3\left(y^4-5y^3(1+\rt)+2y^2(5+11\rt+5\rt^2)-40y\rt(1+\rt)
    -2\rt(5-38\rt+5\rt^2)\right)\cr
    &\ +3x_0^2\left(-2y^4+8y^3(1+\rt)-y^2(15+28\rt+15\rt^2)
    +52y\rt(1+\rt)+18\rt(1-6\rt+\rt^2)\right)\cr
    &\ +6x_0\left(y^4-3y^3(1+\rt)+y^2(5+6\rt+5\rt^2)
    -12y\rt(1+\rt)-8\rt(1-4\rt+\rt^2)\right)
    \Big\)\bigg\}\,,\cr}}
while \poldr\ yields
\eqn\dtgammady{\eqalign{
    {1\over\Gamma_b}&{d\tilde\Gamma\over dy}=(y^2-4\rt)\bigg\{
    x_0^3\Big\(3-y-\rt\Big\)+3x_0^2\Big\(y-2\Big\)\cr
    &+{\lt\ x_0\over m_b^2(1+\rt-y)}\Big\(
    5x_0^2\left(y^2-2y(2-\rt)+6(1-\rt)\right)\cr
    &\ \ +3x_0\left(-5y^2+y(17-5\rt)-4(6-5\rt)\right)
    +3\left(5y^2-10y+12(1-\rt)\right)\Big\)\cr
    &+{\lo\over3m_b^2(1+\rt-y)^2}\Big\(
    3\left(-y^3+2y^2+4y\rt-8\rt\right)\cr
    &\ \
+2x_0^3\left(-y^3+y^2(5+2\rt)+y(-10-11\rt+5\rt^2)+6\rt(5-3\rt)
    \right)\cr
    &\ \ +3x_0^2\left(2y^3-2y^2(4+3\rt)+y(15+22\rt-5\rt^2)
    -24\rt(2-\rt)\right)\cr
    &\ \ +6x_0\left(-y^3+y^2(3+4\rt)-y(5+11\rt)+6\rt(3-\rt)\right)
    \Big\)\bigg\}\,.\cr}}
Here
\eqn\defxzero{
    x_0=1-{\r\over1+\rt-y}\,.}

At this point, we like to mention that there is an elegant
alternative way to obtain the spectra $d\Gamma/dy$ and
$d\tilde\Gamma/dy$. Instead of starting from a triple differential
decay rate, one can construct the operator product expansion for the
decay $\bar B\to\tau+X$, where the neutrino is now part of the final
state $X$. The leading contribution in the expansion is given by a
diagram containing a charm quark--neutrino loop. The discontinuity of
this diagram gives the lepton spectrum. In this approach, the
variable $x_0$ appears in a natural way as the upper limit in the
integration over a Feynman parameter. We have checked that this
appraoch leads to the same results as given above.

Finally, we perform the $y$-integration over the kinematic region
\eqn\yregion{
     2\sqrt{\rt} \le y \le 1 + \rt - \r}
to obtain the total decay rates.  For
$\Gamma$, we find
\eqn\totalrate{\eqalign{{\Gamma\over\Gamma_b}&=
    \sqrt\lambda\bigg\{\Big(1+{\lo\over2m_b^2}\Big)
    \Big\(1-7(\r+\rt)-7(\r^2+\rt^2)+\r^3+\rt^3+
    \r\rt(12-7(\rt+\r))\Big\) \cr &\ +{3\lt\over2m_b^2}
    \Big\(-3+5(\r+\rt)-19(\r^2+\rt^2)+5(\r^3+\rt^3) +7
    \r\rt(4-5(\rt+\r))\Big\)\bigg\}\cr
    &+12\Big(1+{\lo+3\lt\over2m_b^2}\Big)
    \Big\(\r^2\ln{(1+\r-\rt+\sqrt\lambda)^2\over4\r}
    +\rt^2\ln{(1+\rt-\r+\sqrt\lambda)^2\over4\rt}\Big\)\cr
    &-12\Big(1+{\lo+15\lt\over2m_b^2}\Big)\r^2\rt^2\ln
    {(1-\rt-\r+\sqrt\lambda)^2\over4\rt\r}\,,\cr}}
with
\eqn\deflamb{
    \lambda=1-2(\rt+\r)+(\rt-\r)^2\,.}
For $\tilde\Gamma$, we obtain the lengthy expression
\eqn\asym{\eqalign{{\tilde\Gamma\over\Gamma_b}=&
    {1\over6}\((1-\Mt)^2-\r\)\,
    \Big\(-(1-\Mt)^3 (3+15\Mt+5\Mt^2+\Mt^3) \cr
    &\, +\r(1-\Mt)(21+57\Mt+31\Mt^2+11\Mt^3) \cr
    &\, +\r^2(21-15\Mt-5\Mt^2+47\Mt^3)/(1-\Mt) -3\r^3\Big\)\cr
    &-2\r^2 \, (3-3\Mt^4-2\Mt^2 \r) \ln{(1-\Mt)^2\over\r} \cr
    +&{\lo\over12m_b^2} \bigg\{ \((1-\Mt)^2-\r\)\,
    \Big\(-(1-\Mt) (1+\Mt)^3 (3+\Mt^2) \cr
    &\, +\r(1+\Mt) (21-6\Mt+8\Mt^2-2\Mt^3+11\Mt^4)/(1-\Mt) \cr
    &\, +\r^2
    (21-57\Mt+14\Mt^2+42\Mt^3-99\Mt^4+47\Mt^5)/(1-\Mt)^3-3\r^3
    \Big\)\cr
    &-12\r^2\, (3-3\Mt^4-2\Mt^2\r) \ln{(1-\Mt)^2\over\r}
    \bigg\} \cr
    +&{\lt\over 4m_b^2} \bigg\{ \((1-\Mt)^2-\r\)\,
    \Big\( (1-\Mt) (9+27\Mt+70\Mt^2+10 \Mt^3-15\Mt^4-5\Mt^5) \cr
    &\, -\r(15-3\Mt+62\Mt^2-70\Mt^3-45\Mt^4-55\Mt^5)/(1-\Mt) \cr
    &\, +\r^2(57-84\Mt+82\Mt^2+260\Mt^3-235\Mt^4)/(1-\Mt)^2 -15\r^3
    \Big\)\cr
    &-12\r\, (8\Mt^2-8\Mt^4+3\r+4\Mt^2\r-15\Mt^4\r-10\Mt^2\r^2)
    \ln{(1-\Mt)^2\over\r} \bigg\} \,,\cr}}
where
\eqn\defMt{
    \Mt=m_\tau/m_b=\sqrt{\rho_\tau}\,.}

Our results for the differential and total unpolarized decay rates
confirm a very recent calculation of Koyrakh \ref\koyrakh{L. Koyrakh,
TPI-MINN-93/47-T (1993), hep-ph/9311215.}. In the limit $\rt\to 0$,
these results reduce to the expressions given in \bsuv--\mawi. In the
limit $\lo,\lt\to 0$, corresponding to the free quark decay model,
our results agree with ref.~\ref \cpt{J.L. Cortes, X.Y. Pham, and A.
Tounsi, Phys.\ Rev.\ D25 (1982) 188.}. In the same limit, our
expressions for $d\tilde\Gamma/dy$ and $\tilde\Gamma$ agree with
those of ref.~\ref\kal{J. Kalinowski, Phys.\ Lett.\ B245 (1990)
201.}. Finally, for $\Mt=0$ we find $-2d\tilde\Gamma/dy=d\Gamma/dy$
and $-2\tilde\Gamma=\Gamma$, as required.

\newsec{Numerical Analysis}

\subsec{Input parameters}

When we neglect the tiny contribution from $b\to u$ transitions, the
input parameters entering our calculations are the mass of the tau
lepton \ref \mtauexp{ J.Z. Bai {\it et al.}, BES Collaboration,
Phys.\ Rev.\ Lett.\ 69 (1992) 3021; H. Albrecht {\it et al.}, ARGUS
Collaboration, Phys.\ Lett.\ B292 (1992) 221; A. Balest {\it et al.},
CLEO Collaboration, Phys.\ Rev.\ D47 (1993) R3671.},
\eqn\mtaunum{
    m_\tau=1.777\gev,}
the heavy quark masses $m_b$ and $m_c$, the hadronic parameters
$\lambda_1$ and $\lambda_2$, and the quark mixing parameter
$|V_{cb}|$. We stress, however, that not all of these parameters are
independent. In fact, the heavy quark effective theory can be used to
construct a systematic $1/m_Q$ expansion of the masses of hadrons
containing a heavy quark, in which the parameters $\lambda_1$ and
$\lambda_2$ appear at second order. The relevant relations are \FaNe
\eqn\cbmasses{\eqalign{
    m_B &= m_b + \lambar - {\lo+3\lt\over2m_b} + \ldots \,,\cr
    m_D &= m_c + \lambar - {\lo+3\lt\over2m_c} + \ldots \,,\cr}}
where we neglect higher-order power corrections. The parameter
$\bar\Lambda$ can be associated with the effective mass of the light
degrees of freedom in the heavy mesons \AMM. For each set of values
for $\lambar, \lo$, and $\lt$, we solve \cbmasses\ to find the heavy
quark masses that should be used in the results of sect.~2. Hence,
instead of the four parameters $m_b, m_c, \lambda_1, \lambda_2$ we
are left with three paramters $\lambar, \lambda_1, \lambda_2$. It
will turn out that the correlation between the heavy quark masses,
which is imposed by \cbmasses, reduces the theoretical uncertainties
in the results in a significant way.

In our analysis, we use the value of $\lambda_2$ that is obtained
from the known value of the $B-B^*$ mass splitting
[cf.~\deflambda]
\eqn\ltnum{
    \lt={m_{B^*}^2-m_B^2\over4}=0.12\gev^2.}
We expect this value to be accurate up to power corrections of order
$\bar\Lambda/m_b\sim 10\%$. At this point, the parameters $\lambar$
and $\lo$ must still be obtained from nonperturbative calculations,
introducing a certain amount of model dependence. From a QCD sum rule
analysis, one finds that $\lambar$ lies in the range \ref\lob{M.
Neubert, Phys.\ Rev.\ D46 (1993) 1076; see also
ref.~\HQET.}\ref\Baga{E. Bagan, P. Ball, V.M. Braun, and H.G. Dosch,
Phys.\ Lett.\ B278 (1992) 457.}
\eqn\barlam{
    0.45 < \lambar < 0.60\gev\,.}
QCD sum rules have also been used to compute $\lo$ \lob\ref\elet{V.
Eletsky and E. Shuryak, Phys.\ Lett.\ B276 (1992) 191.}\ref\loa{P.
Ball and V.M. Braun, TUM-T31-42/93 (1993), hep-ph/9307291.}, but
these calculations suffer from large uncertainties. There are
theoretical arguments, however, that $\lo$ should be negative (as one
expects, since $-\lo$ is proportional to the kinetic energy of the
heavy quark) \neub, and that its magnitude cannot be too large
\ref\virial{M. Neubert, CERN-TH.7070/93 (1993), hep-ph/9311232, to
appear in Phys.\ Lett.\ B.}. Here we shall use the range
\eqn\numlo{
    0 < -\lo < 0.3\gev^2\,.}

\subsec{${\rm BR}(\bar B\to\tau\,\bar\nu\,X)$}

Let us then present our numerical results. It is convenient to
normalize the total branching ratio ${\rm BR}(\bar
B\to\tau\,\bar\nu\,X)$ to the measured branching fraction into final
states with an electron~\ref \pdg{K. Hikasa {\it et al.}, Particle
Data Group, Phys.\ Rev.\ D45 (1992) S1.},
\eqn\btoe{
    {\rm BR}(\bar B\to e\,\bar\nu\,X)=10.7\pm0.5\ \%\,.}
This eliminates the otherwise significant uncertainties from the
values of $|V_{cb}|^2$ and $m_b^5$. We incorporate the one-loop QCD
corrections, which may be extracted from \ref\pham{Q. Hokim and X.Y.
Pham, Ann.\ Phys.\ 155 (1984) 202.}. Using $\alpha_s(m_b)\approx
0.22$, one finds that the total rates are corrected by multiplicative
factors $\eta_\tau=0.90$ and $\eta_e=0.88$, respectively. What is
relevant for us is the ratio, $\eta_\tau/\eta_e=1.02$, for which the
error due to the choice of scale in the running coupling constant is
very small. From \totalrate, we then obtain the main result of this
section:
\eqn\btotau{
    {\rm BR}(\bar B\to\tau\,\bar\nu\,X) = 2.30\pm 0.25\ \%.}

There are a number of points to be made regarding this result:

$a.$ The $1/m_b^2$ corrections to the free quark decay model reduce
the prediction for the rate by approximately 4\%. For example, if we
take the values $\lambar=0.5$ GeV and $\lambda_1=-0.25$ GeV,
corresponding to the heavy quark masses $m_b=4.8$ GeV and $m_c=1.45$
GeV, the spectator model result of 2.37\% is lowered to 2.28\%. The
fractional decreases in the individual rates are 7.3\% for the tau
channel and 3.8\% for the electron channel.

$b.$ The 11\% uncertainty in our prediction \btotau\ takes into
account the experimental uncertainty in the electronic branching
ratio \btoe, the theoretical uncertainties in \barlam\ and \numlo, as
well as an estimate of higher-order power corrections which we have
not included. The most important of these are the $1/m_c^2$
corrections to the relation \cbmasses\ between $m_D$ and $m_c$. We
estimate the uncertainty related to this effect to be about 2\%. We
emphasize that the largest error of all of these is the {\it
experimental} one, so any improvement in the determination of ${\rm
BR}(\bar B\to e\,\bar\nu\,X)$ will improve the accuracy of our
prediction.

$c.$ Numerically, the main improvement of our calculation over the
spectator model calculation is not in the incorporation of the
nonperturbative $1/m_b^2$ corrections, but in using $m_c$ and $m_b$
as determined by \cbmasses\ rather than treating them as uncorrelated
input parameters. To demonstrate this, let us take for a moment the
central value ${\rm BR}(\bar B\to e\,\bar\nu\,X)=10.7\%$. Then, if we
allow $m_c$ and $m_b$ to vary independently within the ranges
$1.4<m_c<1.5\gev$ and $4.6<m_b<5.0\gev$, we find\foot{It seems to us
that previous results from spectator model calculations ignore the
uncertainty in quark masses and thus significantly underestimate the
theoretical errors.}
${\rm BR}(\bar B\to\tau\,\bar\nu\,X)=2.27\pm0.55\%$. However, if we
use $\lambar$ and $\lo$ as our input parameters and calculate $m_c$
and $m_b$ from \cbmasses, we obtain ${\rm BR}(\bar B\to
\tau\,\bar\nu\,X)=2.30\pm0.09\%$. Accounting for the correlation
between $m_c$ and $m_b$ reduces the error from 24\% to 4\%. Even if
we allow the very conservative range $-0.5<\lo<0.5\gev$ and
$0.4<\lambar<0.7\gev$ (which covers a range for $m_c$ and $m_b$
larger than commonly accepted), we get ${\rm BR}(\bar
B\to\tau\,\bar\nu\,X)=2.23\pm 0.33\%$. This is still more accurate
than if we allow independent variation of $m_c$ and $m_b$ .

$d.$ We have not included the final charmless modes $X_u$ in our
numerical analysis. Since $|V_{ub}/V_{cb}|\sim0.1$~\pdg, their effect
on the inclusive rate for each of the electron and tau channel is of
order 1\%. The effect on the ratio of rates is even smaller and
certainly well below our error in \btotau.

\subsec{The tau polarization}

The polarization of the tau lepton, $A_{\rm
pol}=2\tilde\Gamma/\Gamma$, being a ratio of decay rates, is by
itself subject to much smaller uncertainties than are the rates
themselves. We find that the numerical value for $A_{\rm pol}$ is
rather insensitive to variations of $\lambar$ and $\lo$. Allowing
these parameters to vary within the ranges \barlam\ and \numlo, we
find
\eqn\apolhqet{
    A_{\rm pol}=-0.706\pm0.006\,.}

A few points are in order regarding this result:

$a.$ The $1/m_b^2$ corrections to the free quark decay model reduce
the prediction for the tau polarization by approximately 4\%.

$b.$ The 0.9\% uncertainty in $A_{\rm pol}$ is due almost entirely to
our estimate of higher-order nonperturbative corrections. The
uncertainty from the theoretical errors on $\lambar$ and $\lo$ is
surprisingly small, about 0.1\%.

$c.$ Numerically, the main improvement is again in using $m_c$ and
$m_b$ as determined by \cbmasses. As before, if we allow $m_c$ and
$m_b$ to vary independently, we find $A_{\rm pol}= -0.70\pm 0.04$,
while if we calculate them from $\lambar$ and $\lo$ and allow those
inputs to vary, we find $A_{\rm pol}= -0.7045\pm 0.0005$. Thus, the
correlation between $m_c$ and $m_b$ reduces the error on $A_{\rm
pol}$ from 6\% to 0.1\%.

$d.$ The effect of final charmless hadronic states $X_u$ is included
in the result \apolhqet. The effect, however, is rather small: it
shifts the central value of $A_{\rm pol}$ from $-0.705$ to $-0.706$,
assuming $|V_{ub}/V_{cb}|=0.1$.

$e.$ We have also numerically calculated the tau polarization as a
function of the tau energy, $A_{\rm pol}(y)=-2(d\tilde\Gamma/dy)/
(d\Gamma/dy)$. The $1/m_b^2$ corrections to the spectator results are
small except for a region close to the parton model endpoint
$y_{max}=1-\rho_c+\rho_\tau$. However, in this endpoint region
(corresponding to $|\vec p_\tau|\sim1.6\gev$) our results cannot be
trusted, since the operator product expansion becomes singular, and a
resummation is necessary~\neub\ref\fjmw{A.F. Falk, E. Jenkins, A.V.
Manohar, and M.B. Wise, UCSD/PTH 93-38 (1993), hep-ph/9312306.}
\ref\Fermi{I.I. Bigi, M.A. Shifman, N.G. Uraltsev, and A.I.
Vainshtein, TPI-MINN-93/60-T (1993), hep-ph/9312359.}.

\newsec{An Experimental Bound on $\lo$}

The only quantity concerning inclusive $\bar B\to\tau\,\bar\nu\,X$
decays which has so far been measured is the total branching ratio.
Adding the statistical and systematic errors on the recent ALEPH
measurement~\ref \aleph{I. Tomalin, talk given at the International
Europhysics Conference on High Energy Physics, Marseille (1993).}\ in
quadrature, we will use the value
\eqn\btotauexp{
    {\rm BR}(\bar B\to\tau\,\bar\nu\,X)=2.76\pm 0.63\ \%\,.}
Among the parameters which go into the prediction \btotau, the one
that is subject to the largest theoretical uncertainties is $\lo$.
Using experimental data on inclusive semileptonic $D$ decays (in the
electron channel), one can derive the experimental lower bound
\ref\lusa{M. Luke and M.J. Savage, UCSD/PTH 93-25 (1993),
hep-ph/9308287.}
\eqn\lower{
    \lo\geq\ -0.5\gev^2.}
We find that the semileptonic $B$ decay in the tau channel provides
an upper bound on this quantity. Taking into account the 1$\sigma$
lower bound on the branching ratio \btotauexp, the 1$\sigma$ range
for the branching ratio to electrons \btoe, our range \barlam\ for
$\lambar$, and the uncertainty from higher-order corrections, we
obtain
\eqn\upper{
    \lo\leq\ 0.8\gev^2.}
A lower bound on $\lo$ may also be derived from the same data, but it
is weaker than \lower.

Let us clarify the role of the various input data in deriving the
bound \upper:

$a.$ The upper bound on $\lo$ is sensitive to the lower bound on the
total branching ratio in \btotauexp. If, for example, ${\rm BR}(\bar
B\to\tau\,\bar\nu\,X)=2.76\%$ (the central value of the ALEPH
measurement), then $\lo\leq-0.7\gev^2$ is required, which is
inconsistent with \lower. Consistency of our theoretical analysis
(which is based on the assumption of lepton universality in the
standard model) then suggests that the branching ratio should be at
the lower end of the range quoted by the ALEPH collaboration. This is
also clear if one simply compares our result \btotau\ with the
experimental result \btotauexp.

$b.$ Our bound is also sensitive to the experimental value of ${\rm
BR}(\bar B\to e\,\bar\nu\,X)$. Thus, an improvement in the
experimental data on this mode could lead to very strong constraints
on $\lo$.

$c.$ The bound is weakly dependent on the precise value of $\lambar$.
This is in contrast to the bound derived in \lusa, which is very
sensitive to $\lambar$ through its dependence on $m_c^5$.

$d.$ We emphasize that in deriving the bound \upper, we have allowed
each of the input parameters independently to take their extreme
1$\sigma$ values, so the result we quote is rather conservative.

\newsec{Summary}

Using the operator product expansion and heavy quark effective
theory, we have calculated the inclusive rate for $B$ meson decays
into a tau lepton plus anything, including the effects of the
non-vanishing tau mass and the leading nonperturbative corrections of
order $1/m_b^2$. For the total branching ratio, we find ${\rm
BR}(\bar B\to\tau\,\bar\nu\,X)=2.30\pm 0.25\%$. For the polarization
of the tau lepton, we obtain $A_{\rm pol}=-0.706\pm 0.006$. The
effect of the $1/m_b^2$ corrections on both observables is a small
shift of about 4\% as compared to the free quark decay model.
However, we have emphasized that, from the numerical point of view,
the most significant improvement of our analysis is to implement the
tight correlation between the heavy quark masses $m_b$ and $m_c$,
which is imposed by the structure of the heavy quark expansion. This
correlation reduces the theoretical uncertainties in a very
significant way. The experimental value for the branching ratio
allows us to derive the upper bound $\lo\leq 0.8\gev^2$ on one of the
fundamental parameters of the heavy quark effective theory.

While this paper was in writing, we received a preprint by Balk {\it
et al.}~\ref\bkps{S. Balk, J.G. K\"orner, D. Pirjol, and K.
Schilcher, MZ-TH/93-32 (1993), hep-ph/9312220.}\ with results similar
to ours and to ref.~\koyrakh. Our study of the tau polarization in
sect.~2, and the numerical analysis of sects.~3 and 4, have no
overlap with either of these papers.

\bigbreak
\centerline{{\bf Acknowledgements}}

We thank Ikaros Bigi, Yuval Grossman, and Mark Wise for useful
discussions. AF was supported in part by the United States Department
of Energy under contract DOE-FG03-90ER40546. YN is an incumbent of
the Ruth E.~Recu Career Development chair, and is supported in part
by the Israel Commission for Basic Research, by the United
States--Israel Binational Science Foundation (BSF), and by the
Minerva Foundation.

\listrefs
\end